\documentclass[10pt,twocolumn,letterpaper]{article}

\usepackage[pagenumbers]{iccv} 

\usepackage{algorithm}
\usepackage{algpseudocode}
\usepackage{amsmath}
\usepackage{multirow}
\usepackage{bbm}

\definecolor{Coral}{rgb}{1, 0.47, 0.24}

\definecolor{iccvblue}{rgb}{0.21,0.49,0.74}
\usepackage[pagebackref,breaklinks,colorlinks,allcolors=iccvblue]{hyperref}
\usepackage[accsupp]{axessibility}

\definecolor{darkviolet}{rgb}{0.58, 0.0, 0.83}

\def\paperID{2676} 
\def\confName{ICCV}
\def\confYear{2025}

\title{ObjectGS: Object-aware Scene Reconstruction and Scene Understanding\\ via Gaussian Splatting
}

\author{
{Ruijie Zhu}$^{1,2}$\thanks{The work is done during Ruijie Zhu's internship at Shanghai AI Lab.} ~\quad
{Mulin Yu}$^{2}$ ~\quad 
{Linning Xu}$^{3}$ ~\quad 
{Lihan Jiang}$^{1,2}$ ~\quad 
{Yixuan Li}$^{3}$ ~\quad \\ 
{Tianzhu Zhang}$^{1}$\thanks{Corresponding author.} ~\quad 
{Jiangmiao Pang}$^{2}$ ~\quad 
{Bo Dai}$^{4}$ \\\\
{$^1$ University of Science and Technology of China} \quad
{$^2$ Shanghai Artificial Intelligence Laboratory} \\
{$^3$ The Chinese University of Hong Kong} \quad
{$^4$ The University of Hong Kong}
\\
}

\begin{document}
\maketitle


\begin{abstract}

3D Gaussian Splatting is renowned for its high-fidelity reconstructions and real-time novel view synthesis, yet its lack of semantic understanding limits object-level perception. In this work, we propose ObjectGS, an object-aware framework that unifies 3D scene reconstruction with semantic understanding. Instead of treating the scene as a unified whole, ObjectGS models individual objects as local anchors that generate neural Gaussians and share object IDs, enabling precise object-level reconstruction. During training, we dynamically grow or prune these anchors and optimize their features, while a one-hot ID encoding with a classification loss enforces clear semantic constraints. We show through extensive experiments that ObjectGS not only outperforms state-of-the-art methods on open-vocabulary and panoptic segmentation tasks, but also integrates seamlessly with applications like mesh extraction and scene editing.
Project page: \url{https://ruijiezhu94.github.io/ObjectGS\_page}

\end{abstract}

\section{Introduction}
\label{sec:intro}

3D scene reconstruction and understanding in open-world settings remain challenging yet crucial for applications like embodied AI where robots must recognize and grasp target objects, and film editing, which requires precise 3D object extraction. Recent advances in NeRF~\cite{mildenhall2020nerf, barron2021mip, verbin2022ref} and 3D Gaussian Splatting~\cite{kerbl20233dgs,lu2024scaffold,yu2024mip,ye2025gsplat} have enabled high-quality reconstructions and real-time rendering, but they lack semantic understanding, hindering direct object extraction.
Although 2D Vision Foundation Models~\cite{kirillov2023sam,ren2024grounded} excel at instance segmentation, they fail to maintain 3D consistency across views.

\begin{figure}[t!]
\centering
\includegraphics[width=\linewidth]{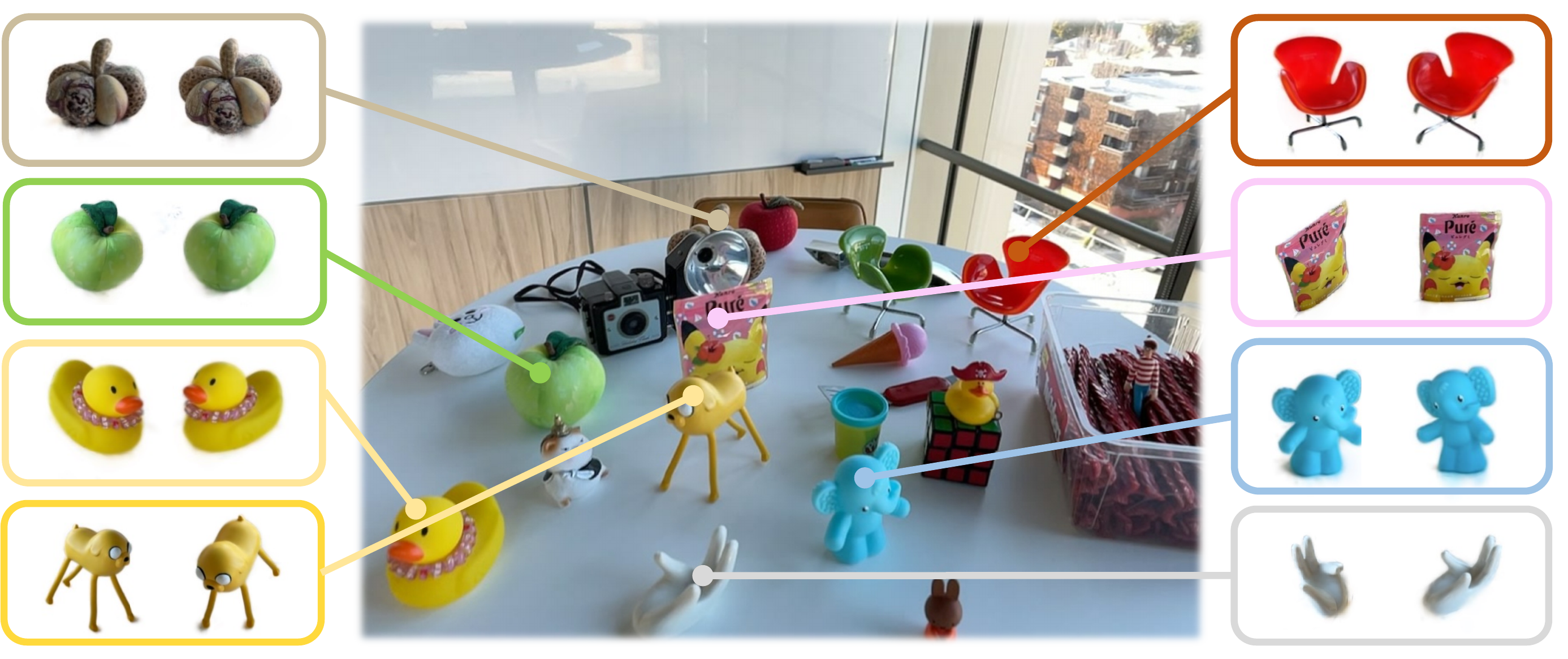}
\caption{In the open-world setting, ObjectGS enables 3D object awareness during reconstruction, allowing it to achieve high-quality scene reconstruction and understanding simultaneously.}
\label{fig:teaser}
\end{figure}

\begin{figure}[!]
    \centering
    \includegraphics[width=\linewidth]{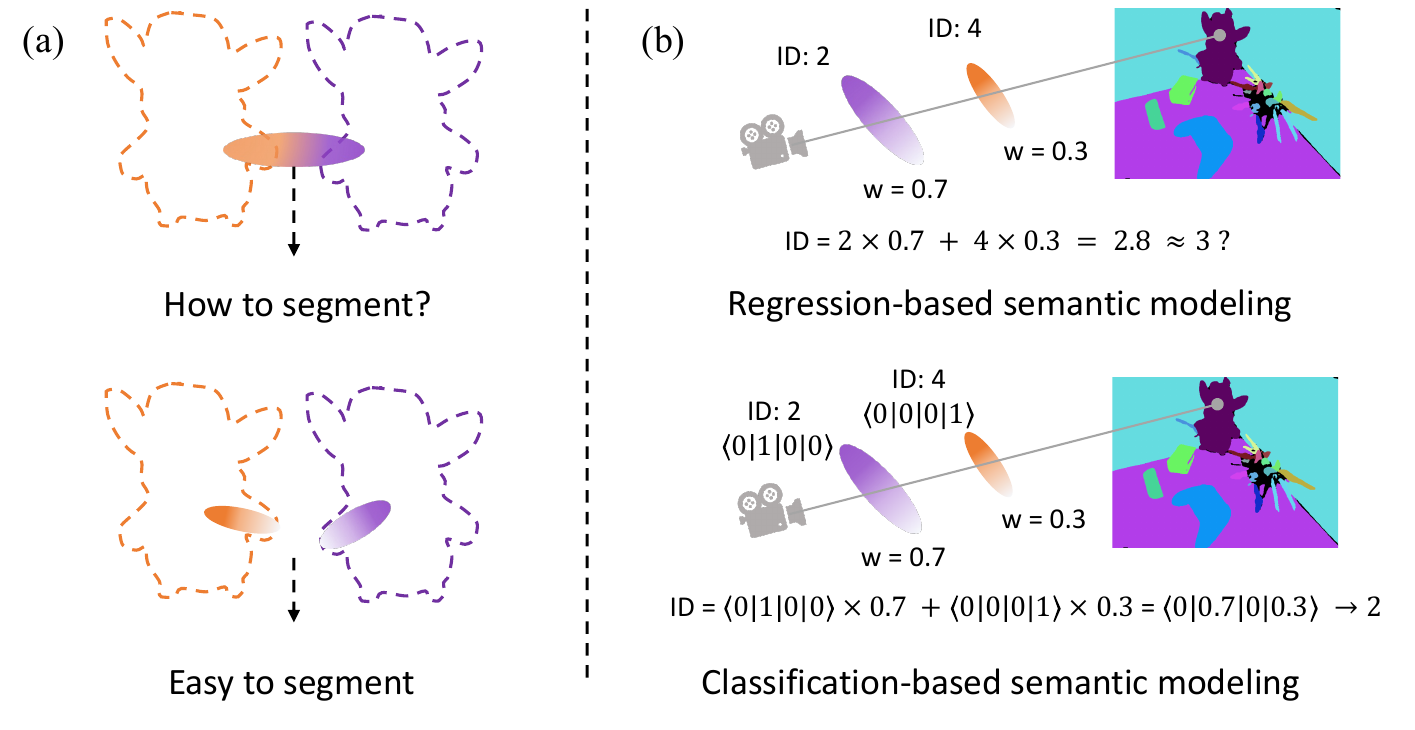}
    \caption{
    (a) Considering semantic information during reconstruction can help better model objects. 
    (b) Existing semantic modeling methods often lead to semantic ambiguity during alpha blending, whereas our classification-based semantic modeling eliminates this problem by independently accumulating the semantics of different objects through ID encoding.
    }
    \label{fig:moti}
\end{figure}

To address this issue, recent approaches~\cite{qin2024langsplat,gaussian_grouping,cen2023saga,wu2024opengaussian,choi2024click} integrate these 2D VFMs into 3DGS frameworks, enabling open-vocabulary segmentation in 3D. Although these methods have achieved 3D instance segmentation in open scenes, we identify two overlooked issues. 
First, some methods~\cite{qin2024langsplat,wu2024opengaussian,cen2023saga} treat 3D reconstruction and segmentation as separate tasks, 
ignoring their inherent interdependence—where precise reconstruction is key to accurate segmentation, and semantic cues help resolve ambiguities.
For instance, as shown in~\cref{fig:moti}(a), it is hard to perform segmentation on a misconstructed Gaussian, but incorporating semantic information during reconstruction can help eliminate such ambiguity.
Second, current approaches~\cite{gaussian_grouping,qin2024langsplat,liang2024supergseg} use \emph{continuous} 3D semantic fields for segmentation, which contradicts the inherently discrete nature of semantic classification and introduces ambiguity during alpha blending.  
As shown in~\cref{fig:moti}(b), regression-based Gaussian semantic features inevitably introduce vagueness in alpha blending.

Building on above analysis, we propose ObjectGS, a Gaussian splatting framework that unifies scene reconstruction and understanding by modeling each object as a collection of Gaussians, as shown in~\cref{fig:teaser}.
Specifically, our method consists of three key components:
(1) \emph{Object ID Labeling and Voting}: 
Leveraging a SAM-based segmentation pipeline, we generate consistent semantic labels across views and employ a majority voting scheme to robustly assign object IDs to the initial scene point cloud, laying a strong foundation for object differentiation.
(2) \emph{Object-aware Neural Gaussian Generation}: 
Building on these object IDs, we introduce a novel strategy inspired by Scaffold-GS~\cite{lu2024scaffold} to generate anchors—minimal modeling units enriched with 3DGS~\cite{kerbl20233dgs} or 2DGS~\cite{huang20242d} primitives—that dynamically grow or prune during reconstruction, ensuring each object's unique features are accurately captured.
(3) \emph{Discrete Gaussian Semantics modeling}: 
To guarantee unambiguous object recognition, we assign each neural Gaussian a fixed one-hot ID encoding based solely on its object ID, a departure from conventional learnable semantics. This discrete representation enables precise 2D splatting and pixel-level object identification, effectively bridging the gap between reconstruction and semantic understanding.

Our main contributions can be summarized as follows: 
\begin{itemize} 
     \item We propose ObjectGS, a novel Gaussian Splatting framework that unifies scene reconstruction and understanding in open-world settings. \item We develop an object-aware training framework that leverages semantic cues to adaptively model objects. \item We introduce a classification-based approach to Gaussian semantics, achieving precise 3D instance segmentation. \item Extensive experiments show that our method outperforms state-of-the-art approaches, while seamlessly supporting scene decomposition and editing.
\end{itemize}


\section{Related Work}
\label{sec:related work}

\paragraph{3D Gaussian Splatting.}
After the tremendous success of Neural Radiance Field (NeRF)~\cite{mildenhall2020nerf} in novel view synthesis and 3D reconstruction, 3D Gaussian Splatting (3DGS)~\cite{kerbl20233dgs} has emerged as the new favorite, gaining significant attention from the research community. 
Compared to NeRF, 3DGS offers explicit scene representation, high-quality reconstruction, and real-time rendering, which presents a broader range of application prospects~\cite{ren2024octree, yu2024gsdf, jiang2025horizon, zhu2024motiongs, lu2024dn4dgs, jiang2025anysplat}. 
Our work is built upon Scaffold-GS~\cite{lu2024scaffold}, with the core idea being the generation of neural Gaussians through anchors, thereby creating a hierarchical scene representation. Furthermore, we extend this framework by modeling the semantics of Gaussians, enabling it to perceive objects in the scene while performing reconstruction. This enhances our ability to simultaneously achieve both 3D scene reconstruction and understanding.

\paragraph{Open-world 2D Segmentation.}
The development of visual foundation models~\cite{vaswani2017attention, caron2021dino,kirillov2023sam,radford2021clip,oquab2023dinov2} has accelerated the application of low-level visual tasks~\cite{cheng2022masked, zhu2023habins, zhu2024scaledepth, song2025depthmaster, liu2024plane2depth, kong2023robodepth, zama2024tricky, zhu2023tiface, jang2023vschh}. Among them, 2D segmentation tasks have gradually started to address general scene segmentation. SAM~\cite{kirillov2023sam} is a milestone in this area, showcasing impressive zero-shot segmentation capabilities in open-world scenarios. It can fulfill specific segmentation needs through flexible prompts, such as points, bounding boxes, or text, and can even perform automatic segmentation without any prompts.
However, SAM does not directly enable cross-frame consistency for video segmentation. Subsequent methods~\cite{cheng2023tracking, ravi2024sam} extended SAM's capabilities to unlock the potential for open-world video segmentation. Despite these advances, using only 2D vision foundation models does not directly solve the problem of 3D scene segmentation. As a result, some early methods~\cite{cen2023samin3d,ying2024omniseg3d,yang2023sam3d, mirzaei2023spinnerf, goel2023isrf, kobayashi2022dff, takmaz2023openmask3d} have begun to explore combining 3D representation models with SAM to lift its capabilities to 3D scene segmentation.

\paragraph{Open-world 3D Scene Understanding.}
With the rise of 3DGS, recent works~\cite{qin2024langsplat,gaussian_grouping,cen2023saga,peng2024gags,choi2024click,lyu2024gaga,liang2024supergseg} have started to combine 3DGS with 2D vision foundation models for open-vocabulary scene understanding.
For example, Langsplat~\cite{qin2024langsplat} combines SAM and CLIP to extract object features and constructs a 3D language field on top of 3DGS using the CLIP features of the objects, enabling open-vocabulary 3D object segmentation. 
Unlike Langsplat, Gaussian Grouping~\cite{gaussian_grouping} directly leverages DEVA~\cite{cheng2023tracking} to extract ID-consistent masks across multiple views, which are then used to supervise the identity features of each Gaussian, enabling efficient 3D segmentation and scene editing. 
By summarizing existing methods, we find that they typically rely on constructing learnable Gaussian semantic features to achieve 3D segmentation. However, due to the inherent sparsity and uniqueness of semantic features, these methods often require additional regularization terms~\cite{gaussian_grouping,peng2024gags,cen2023saga} or contrastive losses~\cite{cen2023saga, choi2024click} to mitigate the ambiguity of Gaussian semantics. 
In contrast, we innovatively propose a new paradigm that constrains deterministic Gaussian semantics to guide object-aware Gaussians to reconstruct their corresponding objects. 


\section{Methodology}
\label{sec:method}

\begin{figure*}[t]
  \centering
  \includegraphics[width=\linewidth]{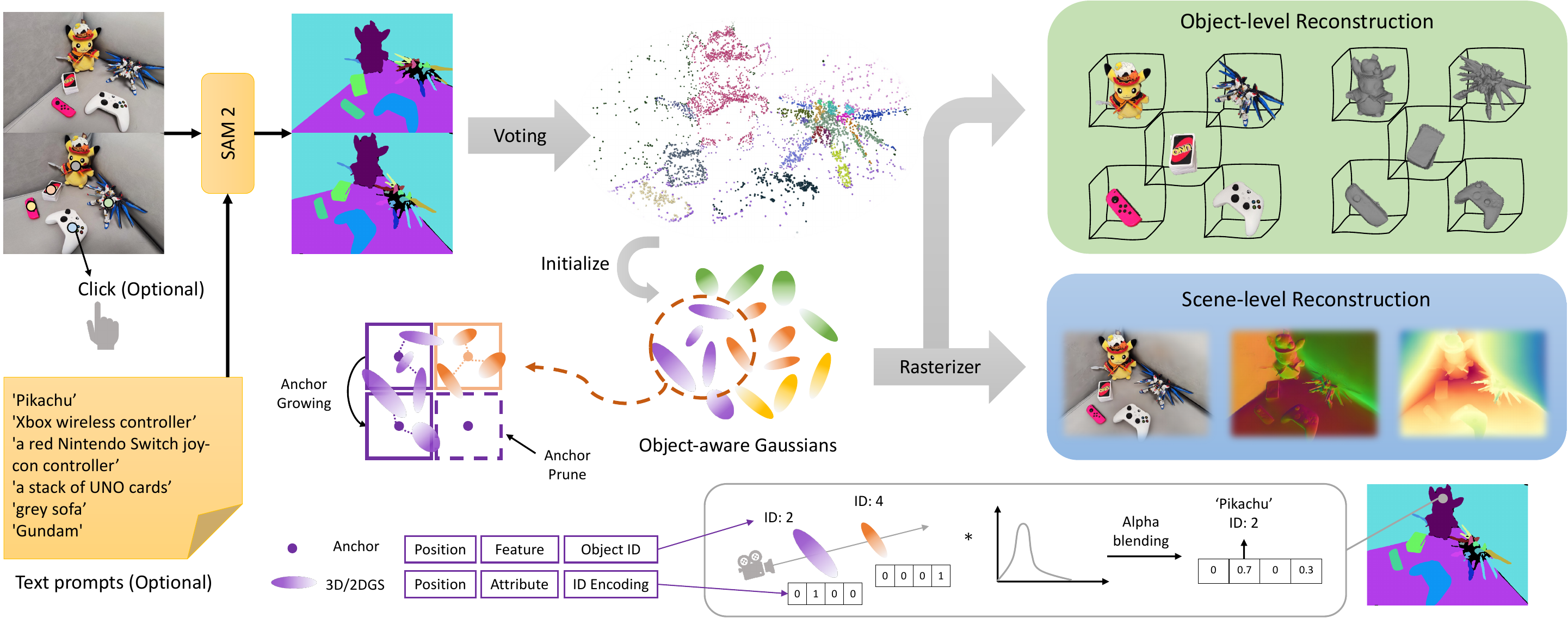}
   \caption{The overall architecture of ObjectGS. We first use a 2D segmentation pipeline to assign object ID and lift it to 3D. Then we initialize the anchors and use them to generate object-aware neural Gaussians. To provide semantic guidance, we model the Gaussian semantics and construct classification-based constraints. As a result, our method enables both object-level and scene-level reconstruction.
   }
   \label{fig:method}
\end{figure*}


The overall architecture of our method is shown in~\cref{fig:method}. In~\cref{sec:initialization}, we first introduce our data preprocessing pipeline, where we extract ID-consistent object masks and use them to initialize the point clouds for different objects. 
In~\cref{sec:neural_gaussian}, we describe how the initialized point cloud generates anchors and their corresponding Gaussians. 
In~\cref{sec:gaussian_semantic}, to enable Gaussian semantic awareness, we model the semantics of the Gaussians and construct classification-based semantic constraints. In~\cref{sec:training_objective}, we introduce the training objectives in our method.

\subsection{Initialization}
\label{sec:initialization}

To consistently lift the semantic information from powerful visual foundation models into 3D, we first extract object masks with consistent IDs across multiple views and then apply a majority voting strategy to assign these masks to each object in 3D space.

\textbf{Object ID Labeling.} Following Gaussian grouping~\cite{gaussian_grouping}, we use DEVA~\cite{cheng2023tracking} to obtain object masks with ID consistency across multiple views. Additionally, to enable open-vocabulary object queries, we also support text and click prompts for selecting specific target objects, with the help of Grounded-SAM~\cite{ren2024grounded}.
Given a sequence of images $\{I_i\}$, we use this pipeline to obtain the ID corresponding to each object in the scene:
\begin{equation}
    \{L_i\} = \text{SAM}(\{I_i\}, \text{Prompts}),
\end{equation}
where the value of $L_i$ indicates the object IDs of pixels in image $I_i$ and the
prompts are optional clicks or texts. Each pixel has an object ID. For some unclassified pixels (no predicted category or invalid value), we uniformly define their ID as 0. Assume that there are $n$ objects in the scene, we assign the object IDs the values $0, 1, 2, ..., n$.

\textbf{Object ID Voting.} The initialization of Gaussian splatting framework relies on point clouds. Therefore, we need to assign the IDs of the object masks to the point cloud. We have already noted that there are some methods, such as~\cite{guo2024samguide}, can segment 3D point clouds aligned with SAM masks. 
However, for the sake of simplicity and ease of use, we design three kinds of voting strategies to quickly initialize the point cloud for different objects.
(1) \emph{Majority Voting.}
Given a sequence of images $\{I_i\}$ with length $N$, the corresponding object ID maps $\{L_i\}$ and a COLMAP point cloud $P_{\text{3D}}$, we first project the point cloud $P_{\text{3D}}$ to 2D views using the camera poses $\{C_i\}$, matching the object ID maps $\{L_i\}$. As a result, each 3D point $P_{i}$ in $P_{\text{3D}}$ has $N$ object ID votes from different views. We use the simple majority voting principle to obtain the object ID of 3D point $P_{i}$, thus deriving the updated point cloud $P_{\text{3D}}$ with object IDs.
(2) \emph{Probability-based Voting.}
Similar to the majority voting, probability-based voting also project point clouds to achieve object-aware voting.
The only difference is that it converts vote counts into probabilities rather than directly taking the majority decision to avoid winner-takes-all situations. 
(3) \emph{Correspondence-based Voting.}
Since the point clouds reconstructed by COLMAP maintain the correspondence between 2D and 3D points, a natural idea is to directly utilize these correspondence as the votes. Therefore, we also try to replace the projecting procedure of the majority voting with the COLMAP correspondence.
The detailed procedure of these three voting strategies are shown in~\cref{sec:voting_supp} of our supplementary.

\subsection{Object-aware Neural Gaussian Generation}
\label{sec:neural_gaussian}

After obtaining the point cloud from the voting process, we use the point clouds corresponding to different objects to initialize anchors, which serve as the carriers for generating and controlling Gaussian primitives. Similar to Scaffold-GS~\cite{lu2024scaffold}, each anchor corresponds to the center of a voxelized grid from the point cloud and carries a local context feature, a scaling factor, and $k$ learnable offsets. Since the initialized anchors may be erroneous or sparse, during training, the anchors adaptively grow and prune in the voxel grid to meet the requirements of scene reconstruction.

\textbf{Object-aware Anchors.} To enable object awareness, we add an object ID to each anchor, which refers to the corresponding object in the scene. During the growing process, anchors replicate their object IDs, while pruning removes the object IDs. This design has two main benefits:
(1) Anchors for the same object can only be generated by anchors with the same object ID, ensuring that newly generated anchors inherit the features of the same object.
(2) Each voxel grid corresponds to at most one anchor and its object ID, ensuring semantic exclusivity and determinism in 3D space.
Through this simple yet effective design, we can generate object-aware anchors as the basic semantic units.

\textbf{Object-aware Neural Gaussians.} For each anchor, we generate $k$ neural Gaussian primitives (3DGS/2DGS)\footnote{In the current implementation, 3DGS and 2DGS primitives cannot coexist in a single model, so only one of them can be chosen at once. Unless otherwise specified, we use the 3DGS primitive by default in this paper.}. The generated Gaussian primitives can be parameterized by their position $\mu$, opacity $\alpha$, color $c$, scale $s$, and quaternion $q$. 
Similar to Scaffold-GS, the Gaussian position can be calculated as:
\begin{equation}
    \{\mu_0, ...,\mu_{k-1} \} = x + \{o_0, ..., o_{k-1} \} \cdot l,
\end{equation}
where $\{o_0, ..., o_{k-1} \}$ are learnable offsets and $x, l$ are the center and scaling factor of the anchor. Other Gaussian attributes such as color $c$ can be computed via an MLP as:
\begin{equation}
    \{c_0, ...,c_{k-1} \} = \text{MLP}(f, \delta, d),
\end{equation}
where $f$ is the anchor feature, $\delta$ and $d$ are the viewing distance and direction between the camera and anchor point.

\subsection{Discrete Gaussian Semantic Modeling}
\label{sec:gaussian_semantic}

In our design, the semantics of the anchors are already modeled using object IDs. A natural idea is to let the generated Gaussian primitives inherit the anchor’s object ID, allowing us to easily achieve semantic modeling in 3D space. However, since the object masks are in 2D space, we need to establish a correspondence between 2D and 3D semantics in order to effectively constrain the Gaussian semantics. After analysis, we identify several approaches for this process:

\begin{figure}[t]
    \centering
    \includegraphics[width=\linewidth]{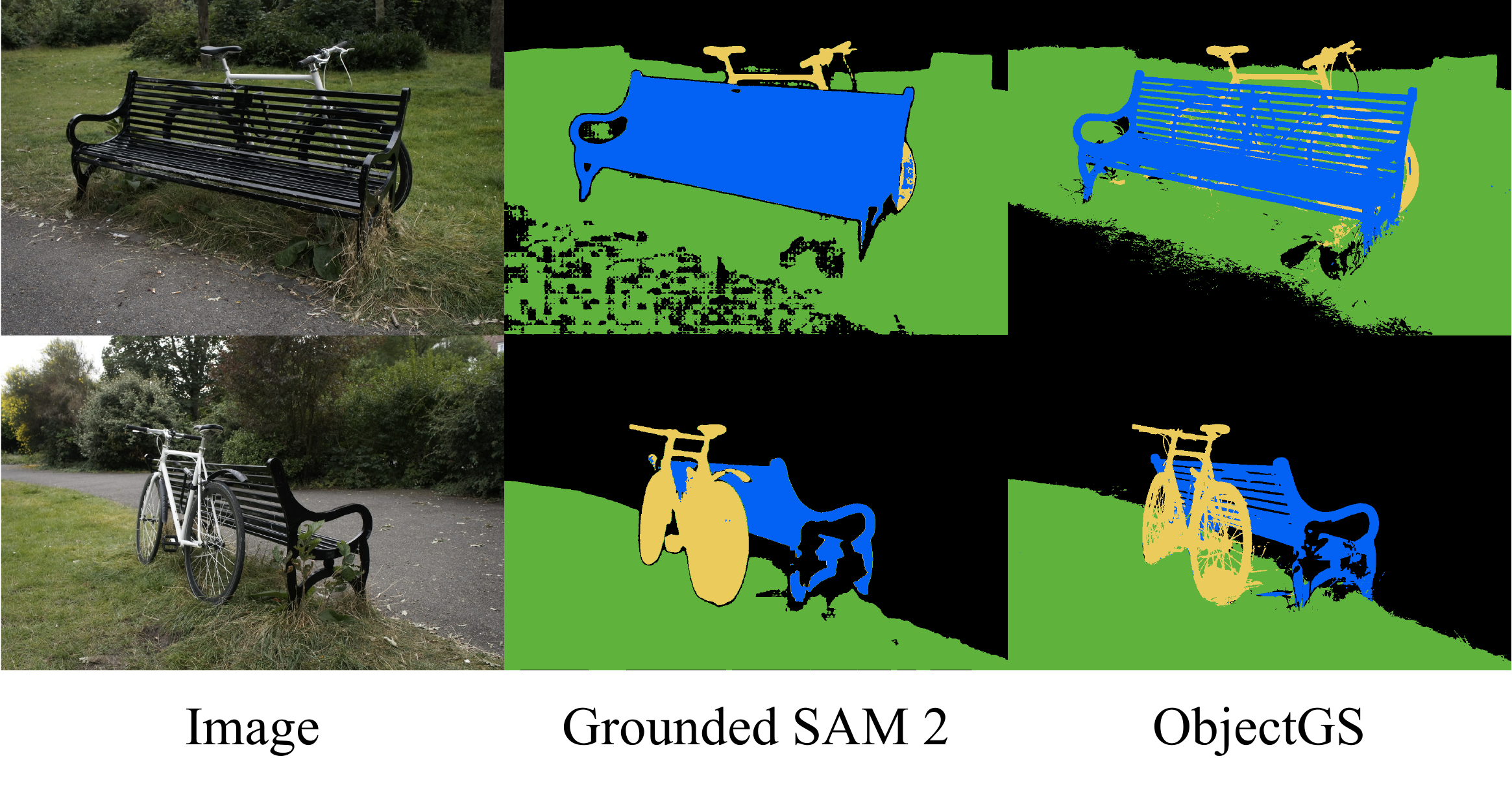}
    \caption{Since 2D segmentation method~\cite{ren2024grounded} don't account for occluded object, it cannot be used to supervise the independent rendering of objects. In contrast, our ObjectGS render semantics in the scene level, which is occlusion-aware.}
    \label{fig:occlussion}
\end{figure}

\textbf{(a) Learnable Gaussian Semantics.}
One simple approach is to follow the color rasterize pipeline to define learnable Gaussian semantic features and optimize them through 2D feature distillation. This approach is widely used in existing methods~\cite{qin2024langsplat,peng2024gags,cen2023saga,chacko2025lifting,liang2024supergseg} of Gaussian semantic modeling. While this approach might seem reasonable, it overlooks a key distinction between the color and semantic attributes of Gaussians: color can be continuous, while semantics are discrete. As mentioned in~\cref{fig:moti}, blending the semantics via alpha blending in this manner could confuse the Gaussian semantics of different categories and introduce ambiguity.

\textbf{(b) Object-independent Constraints.}
To resolve the semantic ambiguity in alpha blending, one possible solution is to query the object IDs and independently render objects. By iterating over all object IDs, we can render the object masks for all objects in the scene and use pseudo ground truths derived in 2D segmentation pipeline for supervision. While this approach may seem feasible, it misses a crucial point as shown in~\cref{fig:occlussion}:  when segmenting objects in 2D images to generate pseudo object mask labels, it don’t account for occluded objects. Similar observation is also included in~\cite{wu2023objectsdf++}. Therefore, this method cannot handle the complexities of occlusion in scenes where objects overlap.

\textbf{(c) One-hot ID Encoding.} To address the above issues, we propose using one-hot ID encoding as the modeling of Gaussian semantics, where the length of the ID encoding is equal to the number of objects in the scene. If there are \(n\) objects, we assign object IDs as \(1, 2, \dots, n\), and for an object with ID \(i\), its one-hot encoding vector \(\mathbf{E}_i\) is defined as:
\begin{equation}
    \mathbf{E}_i = [0, 0, \dots, 1, \dots, 0]. \quad \text{(with 1 in the \(i\)-th dim)}
\end{equation}
Each anchor has an object ID, and all Gaussians generated by the same anchor share the same one-hot ID encoding. 

\textbf{Gaussian Semantic Rendering.} During rendering, alpha blending is performed across the Gaussians along the ray, and the accumulated ID encoding at each pixel is computed as:
\begin{equation}
    \mathbf{P}(\mathbf{x}) = \sum_{k} \alpha_k \cdot T_{k} \cdot \mathbf{E}_{i_k},
\end{equation}
where
\(\alpha_k\) and \(T_{k}\) are the opacity and the accumulated transmittance of the \(k\)-th Gaussian along the ray at pixel \(\mathbf{x}\), 
\(\mathbf{E}_{i_k}\) is the one-hot ID encoding of the \(k\)-th Gaussian with object ID \(i_k\).
\(\mathbf{P}(\mathbf{x})\) is the resulting classification probability vector at pixel \(\mathbf{x}\), which represents the probability of the pixel belonging to each object ID.
Therefore, the predicted object ID of pixels can be derived by taking the index of the maximum classification probability in \(\mathbf{P}(\mathbf{x})\):
\begin{equation}
    \text{ID}(\mathbf{x}) = \arg \max_i (P_i(\mathbf{x})),
\end{equation}
where
\(\text{ID}(\mathbf{x})\) is the predicted object ID for pixel \(\mathbf{x}\),
\(P_i(\mathbf{x})\) is the classification probability of object ID \(i\) at pixel \(\mathbf{x}\) in the vector \(\mathbf{P}(\mathbf{x})\).

\textbf{Gaussian Semantic Loss.}
After deriving the classification probability of the pixel, we can construct a cross entropy loss instead of a L1 loss to constrain the semantics of the Gaussian:
\begin{equation}
    \mathcal{L}_{\text{semantic}} =  - \sum_{\mathbf{x}}\sum_{i=1}^{n} \mathbbm{1}\left({\text{ID}'}(\mathbf{x}) = i\right) \cdot \log \left( P_i(\mathbf{x}) \right),
\end{equation}
where
\(\mathbbm{1}\) is the indicator function, which is 1 if the condition is true and 0 otherwise,
\({\text{ID}'}(\mathbf{x})\) is the ground truth object ID for pixel \(\mathbf{x}\) derived in~\cref{sec:initialization}.
This approach ensures that the Gaussian semantics for different objects do not interfere with each other during alpha blending. Plus, since we only need to perform alpha blending once at the scene level, this method is occlusion-aware and highly efficient.

\textbf{Variable-length Feature Rasterizer.} Although similar semantic modeling methods have been used in NeRF-based approaches~\cite{guo2022neural,wu2022object}, current Gaussian-based methods have not yet adopted this kind of semantic modeling. One possible reason for this is that in the original Gaussian CUDA implementation, Gaussian attributes are of fixed length during rasterization. In contrast, to make the ID encoding length adaptable to scenes with different numbers of objects, we implement a variable-length feature alpha blending. As a result, our Gaussian semantic rendering is both convenient and efficient.
Consequently, we only need parallelly splatting all the Gaussians in the scene at once to obtain the semantics of all corresponding objects.

\subsection{Training Objective}
\label{sec:training_objective}

With the help of our object-aware Neural Gaussians and discrete Gaussian Semantic Modeling, our method is capable of simultaneously performing object-aware scene reconstruction and 3D scene understanding. Our overall training loss can be expressed as:
\begin{equation}
    \mathcal{L} = \mathcal{L}_1 + \lambda_{\text{SSIM}} \mathcal{L}_{\text{SSIM}} + \lambda_{\text{vol}} \mathcal{L}_{\text{vol}} + \lambda_{\text{semantic}} \mathcal{L}_{\text{semantic}},
\end{equation}
where $\mathcal{L}_1$ and $\mathcal{L}_{\text{SSIM}}$ are the appearance loss between rendered images and ground truth images, $\mathcal{L}_{\text{vol}}$ is the volume regularization term in Scaffold-GS~\cite{lu2024scaffold} and $\mathcal{L}_{\text{semantic}}$ is the proposed Gaussian semantic loss.


\section{Experiment}
\label{sec:exp}

\subsection{Experimental Setup}
\label{sec:setup}

\begin{figure*}
    \centering
    \includegraphics[width=\linewidth]{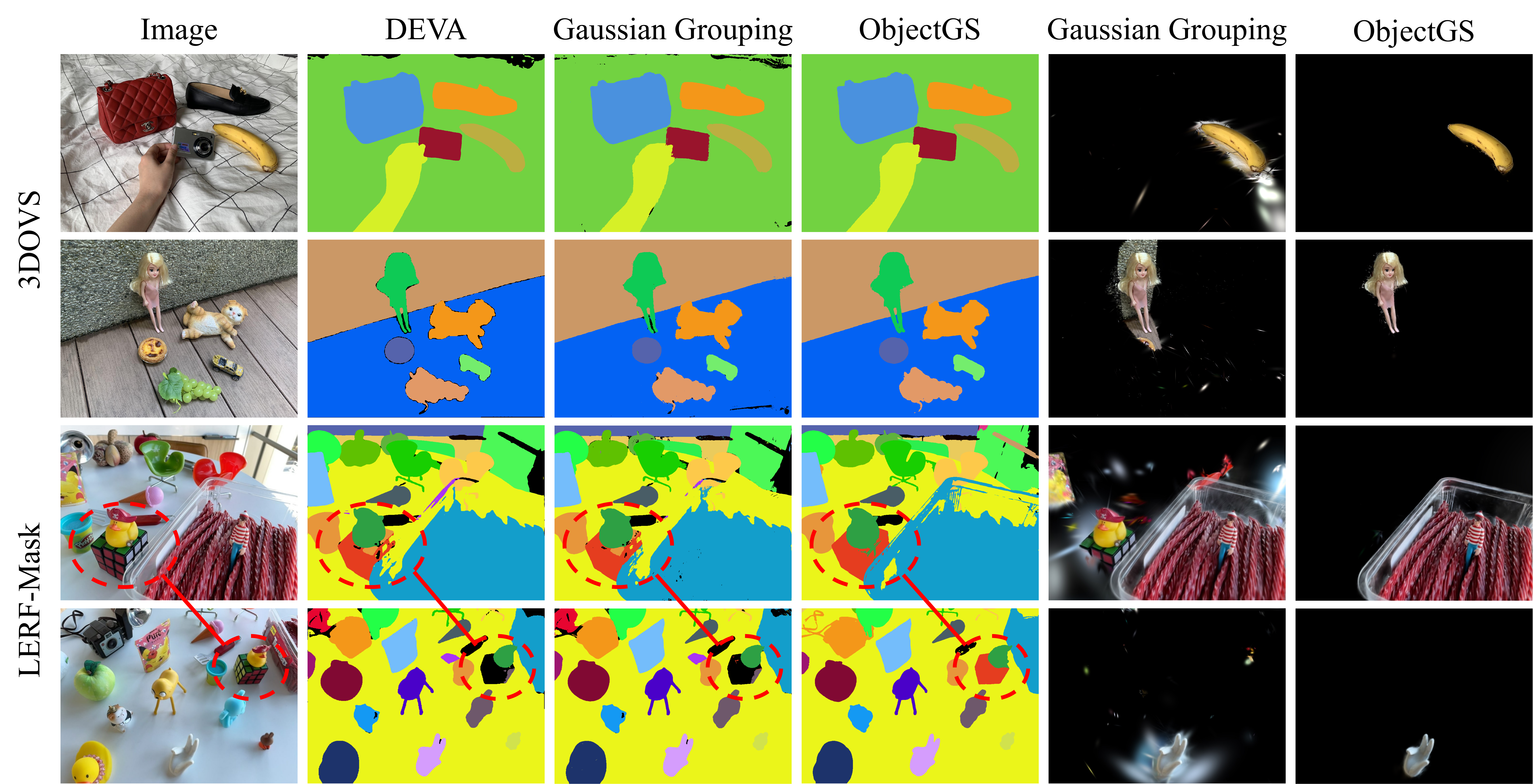}
    \caption{Qualitative comparison of open-vocabulary segmentation and 3D object queries. The red box highlights that our method can achieve multi-view consistent instance segmentation. In 3D object queries, our method has more accurate object segmentation boundaries.}
    \label{fig:ovs}
    \vspace{1em}
    \centering
    \includegraphics[width=\linewidth]{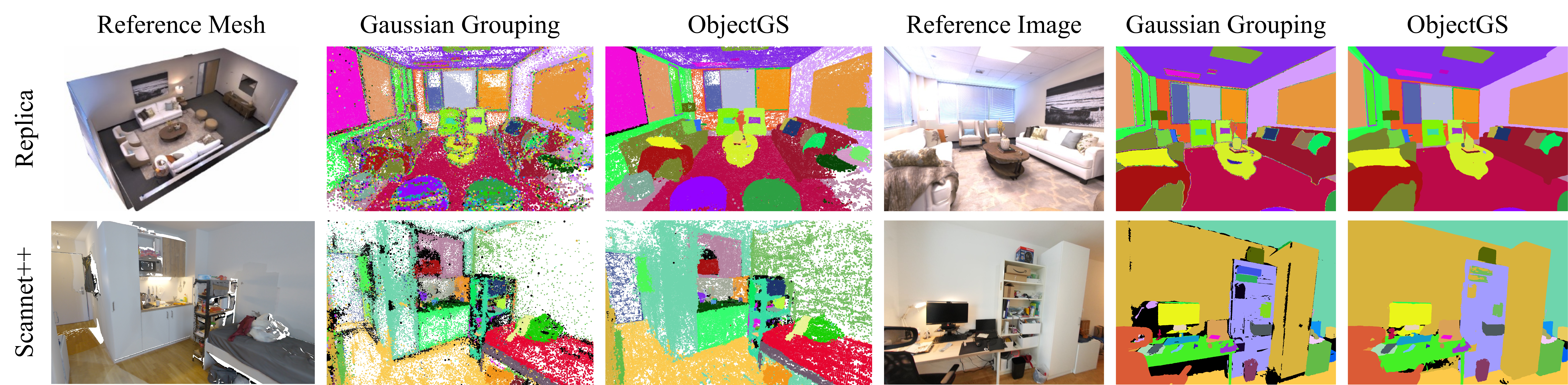}
    \caption{Qualitative comparison of panoptic segmentation. We visualize the segmentation of anchors (ours) and Gaussians (Gaussian Grouping~\cite{gaussian_grouping}) using point clouds, where our results are more consistent and have less noise in 3D space. In 2D instance segmentation, our results have fewer holes and clearer boundaries.}
    \label{fig:pano}
\end{figure*}

\begin{figure*}
    \centering
    \includegraphics[width=\linewidth]{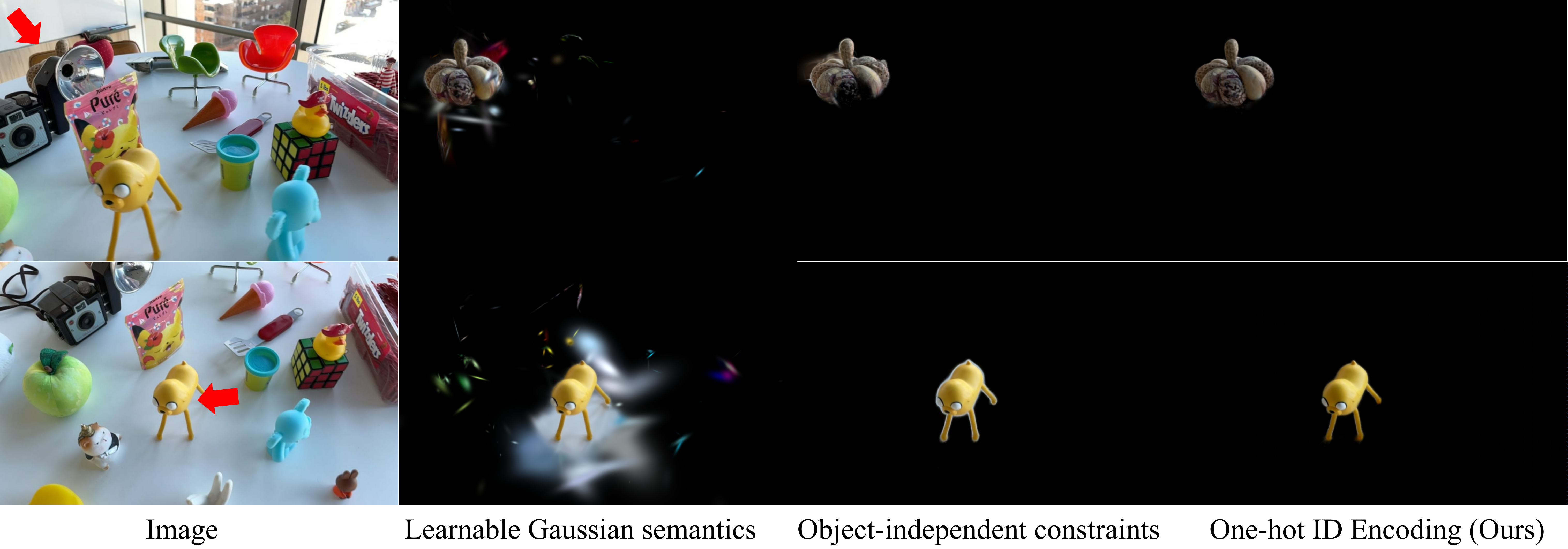}
    \caption{Qualitative comparison of different semantic modeling methods on 3D object query. Learnable Gaussian semantics leads to fuzzy positioning at the object boundary, and the constraint of object independence leads to ineffective object query under occlusion. In contrast, our proposed one-hot ID encoding overcomes both problems and achieves accurate 3D object query.}
    \label{fig:id_encoding}
    \vspace{-1em}
\end{figure*}


\begin{table}[t!]
\caption{Open-vocabulary segmentation results on LERF-Mask dataset. We follow Gaussian Grouping~\cite{gaussian_grouping} to test our method.}
\label{table:lerfmask}
\vspace{-0.5em}
\centering
\resizebox{\linewidth}{!}{
\begin{tabular}{@{}l|cc|cc|cc@{}}
\toprule
\multirow{2}{*}{Model} & \multicolumn{2}{c|}{figurines} & \multicolumn{2}{c|}{ramen}    & \multicolumn{2}{c}{teatime}        \\
                       & mIoU           & mBIoU         & mIoU          & mBIoU         & mIoU          & mBIoU             \\ \midrule
DEVA~\cite{cheng2023tracking}                  & 46.2           & 45.1          & 56.8          & 51.1          & 54.3          & 52.2              \\
LERF~\cite{kerr2023lerf}                   & 33.5           & 30.6          & 28.3          & 14.7          & 49.7          & 42.6          \\
SA3D~\cite{cen2023samin3d}                   & 24.9           & 23.8          & 7.4           & 7.0           & 42.5          & 39.2             \\
LangSplat~\cite{qin2024langsplat}              & 52.8           & 50.5          & 50.4          & 44.7          & 69.5          & 65.6           \\
GS Grouping~\cite{gaussian_grouping}      & 69.7           & 67.9          & 77.0          & 68.7          & 71.7          & 66.1           \\
Gaga~\cite{lyu2024gaga}                   & \textbf{90.7}  & \textbf{89.0} & 64.1          & 61.6          & 69.3          & 66.0           \\
ObjectGS(Ours)         & 88.2           & 85.2          & \textbf{88.0} & \textbf{79.9} & \textbf{88.9} & \textbf{88.6}   \\ 
\bottomrule
\end{tabular}
}
\end{table}

\paragraph{Setting and Datasets.} 
To comprehensively evaluate the performance of our method in open-world 3D scene understanding tasks, we set up two experimental setups: \emph{open-vocabulary segmentation (OVS)} and \emph{panoptic segmentation}. 
For OVS, the goal is to segment target objects in an open scene based on given text prompts. We follow Gaussian Grouping~\cite{gaussian_grouping} to test our method on the LERF-Mask~\cite{kerr2023lerf} and 3DOVS~\cite{liu2023weakly} datasets. 
For panoptic segmentation, we conduct experiments on the Replica~\cite{straub2019replica} and Scannet++~\cite{yeshwanth2023scannet++} datasets. The goal is to perform instance-level segmentation of each object in the scene.

\paragraph{Implementation Details.}
Following the configuration of Scaffold-GS~\cite{lu2024scaffold}, we set the number of Gaussian primitives per anchor to \(k = 10\) in all our experiments. We use GSplat~\cite{ye2025gsplat} to render the Gaussian primitives. The key difference is that we extend the dimensionality of the Gaussian color attributes from 3 to \(N + 3\), where \(N\) is the number of objects in the scene, defined when assigning object IDs. This makes the semantic rendering of Gaussians efficient. In our experiments, the loss weight \(\lambda_{\text{SSIM}}\) is set to 0.2. For the 3DGS version, we set the volume weight \(\lambda_{\text{vol}}\) to 0.0002 on the 3DOVS dataset, 0.00005 on the LERF-Mask dataset, and 0.00002 on the Replica and ScanNet datasets. For the 2DGS version, we reduce the \(\lambda_{\text{vol}}\) weight by half compared to the 3DGS version.
We train each scene for 30,000 iterations on a single A800 GPU. In the case of the LERF-Mask dataset, we set $\lambda_{\text{semantic}}$ to 0.01, while for other scenes, we set $\lambda_{\text{semantic}}$ to 0.1. 

\subsection{Comparison with the State-of-the-arts}

\begin{table}[t!]
\caption{Panoptic segmentation results on Replica and ScanNet++ datasets. We randomly select 7 scenes in Scannet++ for test.}
\label{table:pano}
\vspace{-0.5em}
\centering
\resizebox{\linewidth}{!}{
\begin{tabular}{@{}l|c|ccc|ccc@{}}
\toprule
Model             & Dataset                    & PSNR           & SSIM            & LPIPS           & IoU            & Dice           & Acc            \\ \midrule
Gaussian Grouping & \multirow{2}{*}{Replica}   & 39.52          & 0.9785          & 0.0548          & 83.36          & 91.84          & 94.70          \\
ObjectGS(Ours)    &                            & \textbf{40.26} & \textbf{0.9842} & \textbf{0.0280} & \textbf{88.39} & \textbf{92.39} & \textbf{95.65} \\ \midrule
Gaussian Grouping & \multirow{2}{*}{Scannet++} & 28.35          & 0.9296          & 0.1641          & 89.82          & 92.91          & 98.44          \\
ObjectGS(Ours)    &                            & \textbf{30.24} & \textbf{0.9327} & \textbf{0.1488} & \textbf{95.38} & \textbf{97.48} & \textbf{99.07} \\ \bottomrule
\end{tabular}
}
\end{table}

\begin{table}[t!]
    \centering
    \caption{Comparison of 3D Instance Segmentation on ScanNet++}
    \label{table:3dseg}
    \vspace{-0.5em}
    \resizebox{\linewidth}{!}{
    \begin{tabular}{lcccc}
        \toprule
        Method & Chamfer Distance $\downarrow$ & Precision $\uparrow$ & Recall $\uparrow$ & F1 Score $\uparrow$ \\
        \midrule
        Gaussian Grouping & 0.1472 & 35.9\% & 66.5\% & 41.6\% \\
        ObjectGS(Ours) & \textbf{0.1132} & \textbf{36.3\%} & \textbf{86.1\%} & \textbf{43.4\%} \\
        \bottomrule
    \end{tabular}
    }
\end{table}

\begin{table}[t!]
\caption{Open-vocabulary segmentation results on 3DOVS dataset. We report IoU metric to compare with other methods. }
\label{table:3dovs}
\centering
\resizebox{\linewidth}{!}{
\begin{tabular}{@{}l|ccccc|c@{}}
\toprule
Method            & bed           & bench         & room          & lawn          & sofa          & MEAN          \\ \midrule
LSEG~\cite{li2022language}            & 56.0   & 6.0           & 19.2          & 4.5           & 17.5          & 20.6                   \\
OVSeg~\cite{liang2023open}         & 79.8      & 88.9          & 71.4          & 66.1          & 81.2          & 77.5                  \\
LERF~\cite{kerr2023lerf}        & 73.5        & 53.2          & 46.6          & 27.0          & 73.7          & 54.8                  \\
3DOVS~\cite{liu2023weakly}         & 89.5    & 89.3          & 92.8          & 74.0          & 88.2          & 86.8                    \\
Langsplat~\cite{qin2024langsplat}         & 77.8          & 77.3          & 58.4          & 90.9          & 60.2          & 73.0          \\
Gaussian Grouping~\cite{gaussian_grouping} & 64.5          & 95.6          & 96.4          & 97.0          & 91.3          & 89.1          \\
SAGA~\cite{cen2023saga}              & 97.4          & 95.4          & \textbf{96.8} & 96.6          & 93.5          & 96.0          \\
LBG~\cite{chacko2025lifting}               & 97.7          & 96.3          & 95.9          & \textbf{97.3} & 87.4          & 94.9          \\
ObjectGS(Ours)    & \textbf{98.0} & \textbf{96.4} & 95.1          & 97.2          & \textbf{95.4} & \textbf{96.4} \\ \bottomrule
\end{tabular}
}
\end{table}

We provide more visualization results (\cref{fig:3dovs,fig:lerf-mask,fig:replica_pc,fig:scannetpp_pc,fig:replica,fig:scannetpp}) in the supplementary materials.

\textbf{Open-Vocabulary Segmentation (OVS).}
~\cref{table:lerfmask,,table:3dovs} show the performance when using text prompts to query objects from LERF-Mask and 3DOVS datasets. 
We use IoU (Intersection over Union) and Boundary IoU as our evaluation metrics. Our method significantly outperforms other approaches on both two OVS benchmarks, demonstrating the superiority of our unique framework design. 
We also provide a qualitative comparison in~\cref{fig:ovs,,fig:3dovs,,fig:lerf-mask} against state-of-the-art methods, where our approach fills most of the mask holes automatically and achieves more precise object segmentation. 
Notably, benefiting from the object id design bound to the anchor, our method can query the target object more accurately and conveniently than Gaussian grouping~\cite{gaussian_grouping} without any post-processing.
Besides, in addition to supporting text-based object queries, our method also supports click-based object queries, which is similar to the implementation in SAGA~\cite{cen2023saga} and Click Gaussian~\cite{choi2024click}.

\textbf{Panoptic Segmentation.}
~\cref{table:pano} demonstrates the performance when lifting the 2D object masks to 3D from Replica and ScanNet++ datasets.
We use IoU, Dice coefficient, and Pixel Accuracy as our evaluation metrics.
Experimental results show that our method outperforms Gaussian Grouping~\cite{gaussian_grouping} in both reconstruction accuracy and segmentation precision. 
We provide visualizations of the segmentation results in~\cref{fig:pano,fig:replica,fig:scannetpp}, demonstrating that our approach produces fewer holes and captures more accurate details. 
More importantly, we visualize the semantics of the point cloud derived from anchors and Gaussians to compare 3D instance segmentation performance. As shown in~\cref{fig:pano,fig:replica_pc,fig:scannetpp_pc}, the point cloud produced by our method exhibit consistent semantics in 3D, whereas Gaussian Grouping~\cite{gaussian_grouping} struggles to maintain this 3D semantic consistency. 
To validate the model performance in 3D Segmentation, we design an evaluation on ScanNet++ dataset: for each instance, we compute Chamfer Distance and F1 score between the reconstructed and ground‑truth point clouds, counting a predicted point as a true positive if it lies within $\tau$=0.02m of any ground‑truth point. As shown in~\Cref{table:3dseg}, our model outperforms GaussianGrouping in all four metrics.
We attribute this to our discrete Gaussian semantic modeling, which ensures that the semantics of different objects remain distinct and unaffected by one another.

\subsection{Ablation Study}
\label{sec:ablation}

To comprehensively demonstrate the effectiveness of each component of our method, we design a series of ablation studies on the LERF-Mask and Replica datasets.

\begin{table}[t!]
\caption{Ablation of Gaussian semantic modeling on figurines scene of LERF-Mask dataset.}
\label{table:semantic_modeling}
\vspace{-0.5em}
\centering
\resizebox{\linewidth}{!}{
\begin{tabular}{@{}l|cc|ccc@{}}
\toprule
Setting & mIoU  & mBIoU & PSNR  & SSIM   & LPIPS  \\ \midrule
Learnable Gaussian Semantics   & 69.57 & 67.86 & 25.67 & 0.8876 & 0.1584 \\
Object-independent constraints            & 37.48 & 35.21 & 25.14 & 0.8911 & 0.1741 \\
One-hot ID Encoding (Ours) & \textbf{88.19} & \textbf{85.22} & \textbf{26.75} & \textbf{0.9134} & \textbf{0.1386} \\ \bottomrule
\end{tabular}
}
\end{table}

\paragraph{Gaussian Semantic Modeling.}
We conduct an ablation study on the figurines scene of the LERF-mask dataset to demonstrate the superiority of our unique semantic modeling approach. Specifically, we compare our method with other semantic modeling methods in~\cref{sec:gaussian_semantic}.
As shown in~\cref{table:semantic_modeling}, our proposed One-hot ID Encoding method significantly outperforms both alternatives, highlighting the effectiveness of our approach. We also visualize the results of rendering individual target objects for each method, as shown in~\cref{fig:id_encoding}. Due to the ambiguity introduced by learnable Gaussian semantics, it struggles to accurately segment the boundaries of objects. Although object-independent constraints can accurately segment the boundaries of objects, it is difficult to solve the rendering of objects in the case of occlusion. In contrast, our method combines the strengths of both approaches, enabling accurate object queries and robust scene decomposition simultaneously.

\begin{figure}[t]
    \centering
    \includegraphics[width=\linewidth]{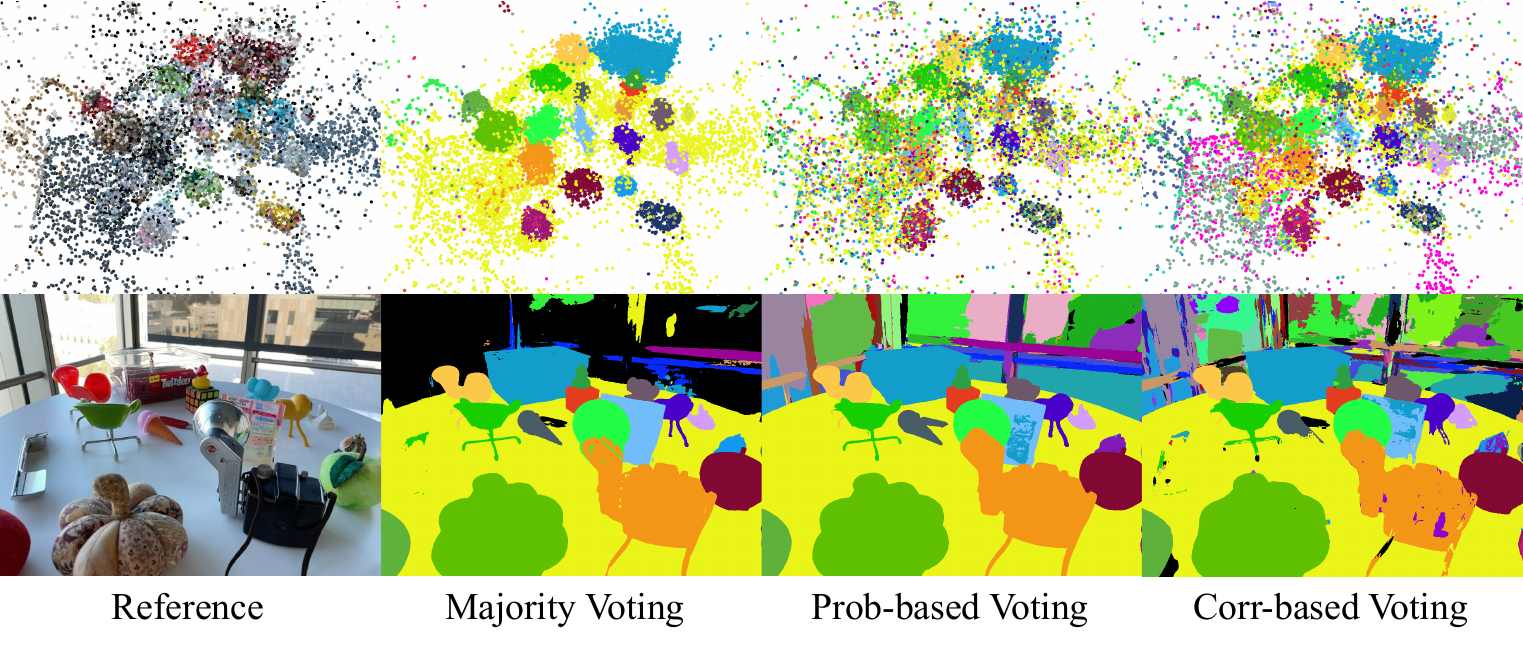}
    \vspace{-2em}
    \caption{Ablation on different point cloud label initializations. The majority voting strategy is more robust in the foreground regions, while the probability-based and correspondence-based voting strategies show greater robustness in the background regions.}
    \label{fig:voting}

\end{figure}

\paragraph{Object ID Voting Strategy}
Since the object ID prediction itself is prone to errors, lifting these predictions to the 3D point cloud inevitably introduces mislabeled points. To validate the robustness of our method, we design and compare three kinds of voting strategy to lift the object masks to 3D. 
As shown in~\cref{fig:voting,table:voting}, though the probability-based and correspondence-based strategy offer relatively more robust results in background regions, they produce suboptimal results when rendering foreground objects compared with the majority voting strategy. We argue that it is due to the grow-and-prune mechanism of our anchors, our method can naturally correct some of these mislabeled points over time. As a result, the simple majority voting strategy proves sufficient for most of the tested scenes.

\begin{table}[t]
\caption{Ablation on object ID voting strategy on figurines scene of LERF-Mask dataset.}
\label{table:voting}
\vspace{-0.5em}
\centering
\resizebox{0.85\linewidth}{!}{
\begin{tabular}{@{}l|cc|ccc@{}}
\toprule
                  & mIoU  & mBIoU & PSNR  & SSIM   & LPIPS  \\ \midrule
Prob-based voting & 84.46 & 81.46 & 25.69 & 0.9019 & 0.1586 \\
Corr-based voting & 59.67 & 57.50 & 26.13 & 0.9031 & 0.1539 \\
Majority voting   & \textbf{88.19} & \textbf{85.22} & \textbf{26.75} & \textbf{0.9134} & \textbf{0.1386} \\ \bottomrule
\end{tabular}
}
\end{table}

\begin{table}[t]
\caption{Ablation of Gaussian semantic loss weights on Replica.}
\label{table:semantic_loss}
\vspace{-0.5em}
\centering
\resizebox{0.8\linewidth}{!}{
\begin{tabular}{@{}c|ccc|ccc@{}}
\toprule
                   $\lambda_{\text{semantic}}$& Acc            & Dice           & mIoU           & PSNR           & SSIM            & LPIPS           \\ \midrule
0.00 & 0.00           & 0.00           & 0.00           & 40.19 & 0.9823          & 0.0288 \\
0.01     & 94.75          & 90.70          & 86.15          & \textbf{40.35}         & 0.9829          & \textbf{0.0273}          \\
0.10     & \textbf{95.65} & \textbf{92.39} & \textbf{88.39} & 40.26          & \textbf{0.9842} & 0.0280          \\
1.00      & 94.42          & 90.98          & 86.67          & 35.43          & 0.9664          & 0.0866          \\ \bottomrule
\end{tabular}
}
\end{table}

\vspace{-1em}

\paragraph{Gaussian Semantic loss.}
To evaluate the effectiveness of semantic constraints, we test our method on the Replica dataset with different weights of semantic loss, as shown in~\cref{table:semantic_loss}. The results show that with a properly chosen loss weight, supervising Gaussian semantics helps improve both scene reconstruction and scene understanding.

\subsection{Application}
\label{sec:application}

Our explicit object-aware Gaussian representation enables several downstream applications post-training. We demonstrate two examples, as shown in our demo video:

\textbf{Object Mesh Extraction.} 
For object mesh extraction, we leverage our 2DGS-based variant. Specifically, we replace 3DGS primitives with 2DGS~\cite{huang20242d} because 2DGS typically better represents object surfaces. Once the scene is reconstructed, we can select target objects using either text prompts or click prompts. Since the object ID is directly bound to the anchor, we can use the anchors with the corresponding ID to generate the 2DGS model of the target object. We then apply TSDF Fusion, as suggested by 2DGS, to export the target object's mesh.  

\textbf{Scene Editing.}
For scene editing, we adopt strategies similar to Gaussian Grouping~\cite{gaussian_grouping}. Moreover, our method can more conveniently select the editing object, without calling the classifier. For example, object removal can be easily achieved by deleting the anchors associated with the target object's ID. To recolor objects, we directly modify the color attributes of the associated Gaussians.


\section{Limitation}
\label{sec:limitation}

Although our method achieves robust open-world scene reconstruction and understanding in our test scenarios, some limitations still exist. Like existing approaches, we rely on 2D segmentation models~\cite{ren2024grounded,cheng2023tracking} to extract object masks. Therefore, when the segmentation model is unavailable or produces severely erroneous outputs, our method may fail. However, our approach is not merely a direct fitting of the 2D segmentation results. In our experimental results (\ie \cref{fig:ovs,fig:pano}), our method demonstrates fewer holes and more 3D-consistent results than the ground truth, indicating that our method can leverage scene geometry to infer unclassified semantics or correct misclassified semantics.

\section{Conclusion}
\label{sec:conclusion}

We propose ObjectGS, an object-aware Gaussian splatting framework for open-world 3D scene reconstruction and 3D scene understanding. Unlike existing methods that distill Gaussian semantics, we optimize object-aware anchors to adjust Gaussian semantics. This design enables our method to perceive objects during reconstruction and adaptively build Gaussian representations based on the needs of individual objects. Furthermore, unlike existing approaches that optimize learnable Gaussian semantics, we model discrete Gaussian semantics and introduce a classification loss. This way ensures that Gaussians from different categories do not interfere during  rendering. 
Finally, we demonstrate the extensibility of our method through its applications in object mesh extraction and scene editing, showcasing its versatility in downstream tasks.  
\clearpage
\section*{Acknowledgement}
\label{sec:acknowledgement}
The work was supported by National Key Research and Development Program of China (2024YFB3909902), National Nature Science Foundation of China (62121002), Youth Innovation Promotion Association of CAS, and HKU Startup Fund.
{
    \small
    \bibliographystyle{ieeenat_fullname}
    \bibliography{main}
}


\clearpage
\setcounter{page}{1}
\maketitlesupplementary

\section{Training Overhead}
\label{sec:overhead}

\Cref{table:trainingtime} compares training time, FPS, and GPU memory across different instance counts. Even with about 100 instances, overhead remains minimal with efficient parallel rasterizer. 
Notably, since our one-hot ID encoding is not learnable parameters, it will not significantly increase training overhead.
Meanwhile, we can optionally encode only a subset of target instances or leverage category hierarchies, avoiding the waste and inflexibility of fixed‑length representations under long‑tailed distributions. 
Therefore, in real applications, our method is both more flexible and scalable.

\begin{table*}[htbp]
    \centering
    \caption{Training time, FPS, and GPU memory comparison}
    \label{table:trainingtime}
    \vspace{-0.5em}
    \resizebox{0.8\linewidth}{!}{
    \begin{tabular}{l|c|cc|cc|cc}
        \toprule
        \multirow{2}{*}{Scene} & \multirow{2}{*}{\#Objects} & \multicolumn{2}{c|}{Training time}    & \multicolumn{2}{c|}{FPS} & \multicolumn{2}{c}{GPU memory} \\
         &  & GS Grouping & Ours & GS Grouping & Ours & GS Grouping & Ours \\
        \midrule
        bed (3DOVS)   & 7   & 94 min & 72 min &  100 & 80 & $\sim$15G  &  $\sim$10G \\
        sofa (3DOVS)) & 24  & 55 min  & 31 min & 110  & 90 &  $\sim$18G &  $\sim$12G \\
        1ada (ScanNet++)  & 63  & 68 min & 69 min & 90  &  50 & $\sim$40G  &  $\sim$35G \\
        3e8b (ScanNet++)  & 80  & 71 min  & 113 min & 80  & 40  & $\sim$40G  & $\sim$45G \\
        0d2e (ScanNet++)  & 90  & 73 min & 112 min & 80  & 40 & $\sim$40G  & $\sim$45G \\
        \bottomrule
    \end{tabular}
    }
    \label{tab:training_time}
\end{table*}

\section{Voting Algorithm}
\label{sec:voting_supp}

We provide the pseudo code of~\cref{alg:major_voting,alg:prob_voting,alg:corr_voting} to clearly demonstrate the proposed voting strategies.

\section{More Visualization}
\label{sec:visualization}

\begin{figure*}
    \centering
    \includegraphics[width=\linewidth]{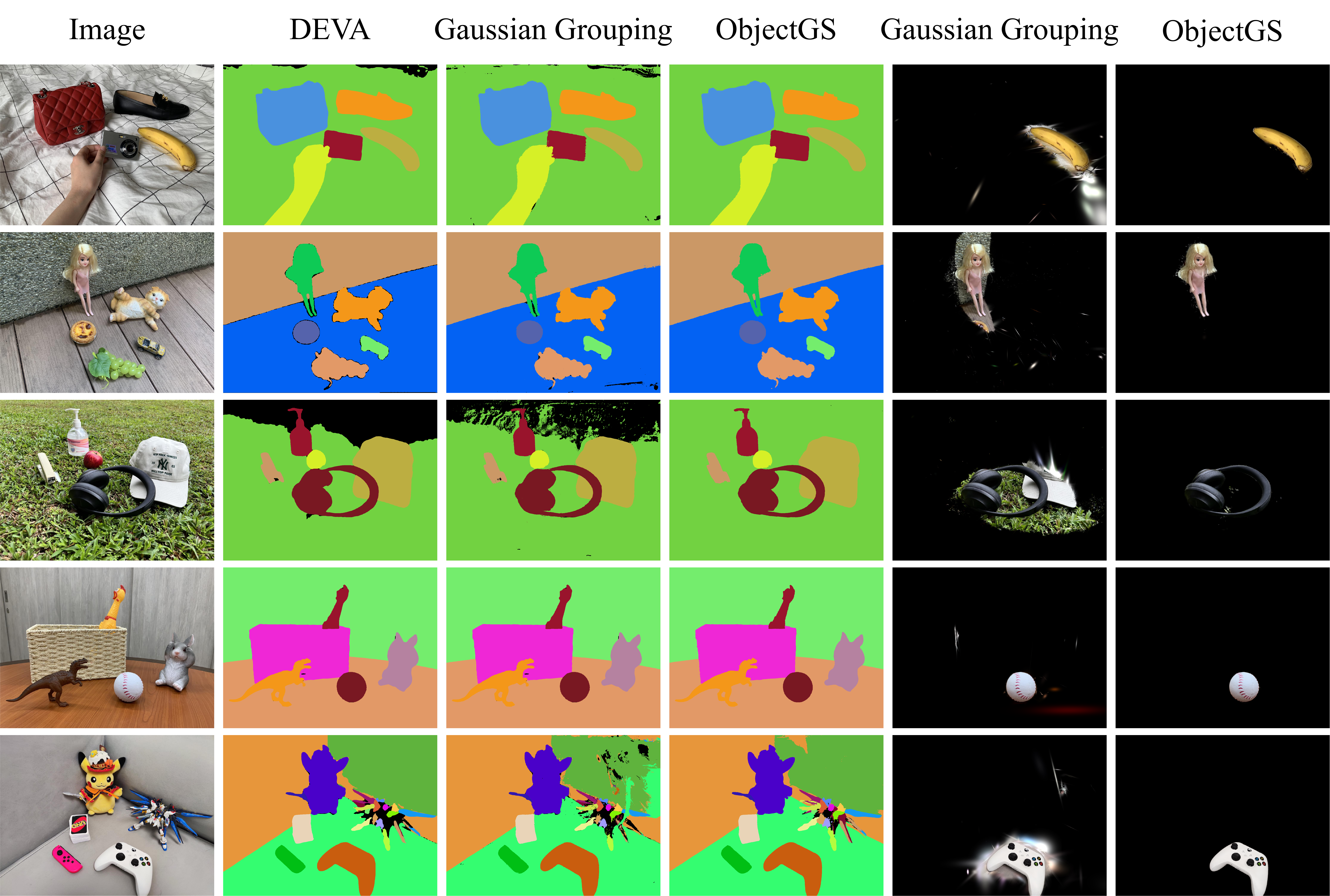}
    \caption{Qualitative comparison of open vocabulary segmentation and 3D object query on the 3DOVS dataset.}
    \label{fig:3dovs}
\end{figure*}

\begin{figure*}
    \centering
    \includegraphics[width=\linewidth]{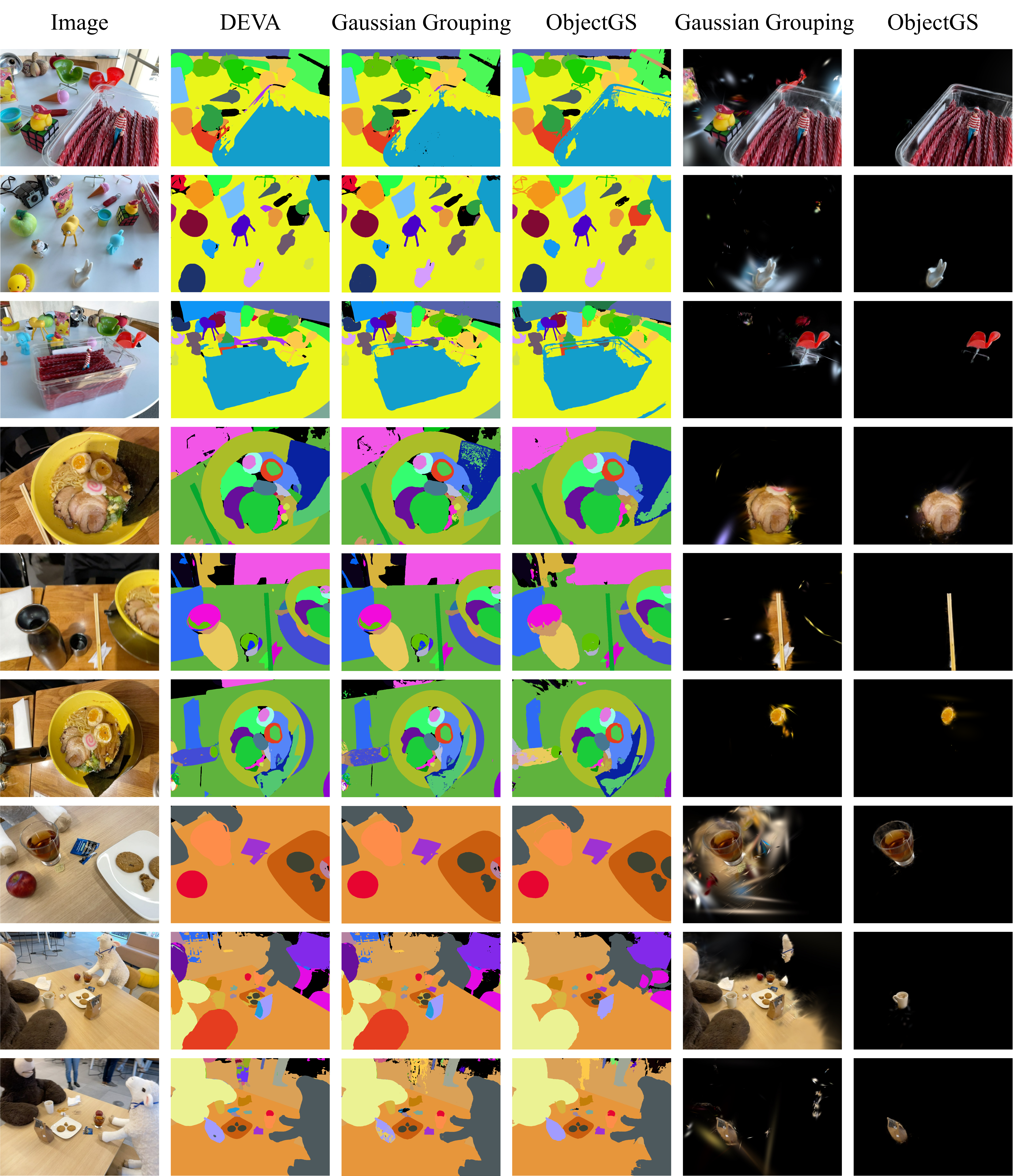}
    \caption{Qualitative comparison of open vocabulary segmentation and 3D object query on the LERF-Mask dataset.}
    \label{fig:lerf-mask}
\end{figure*}

\begin{figure*}
    \centering
    \includegraphics[width=\linewidth]{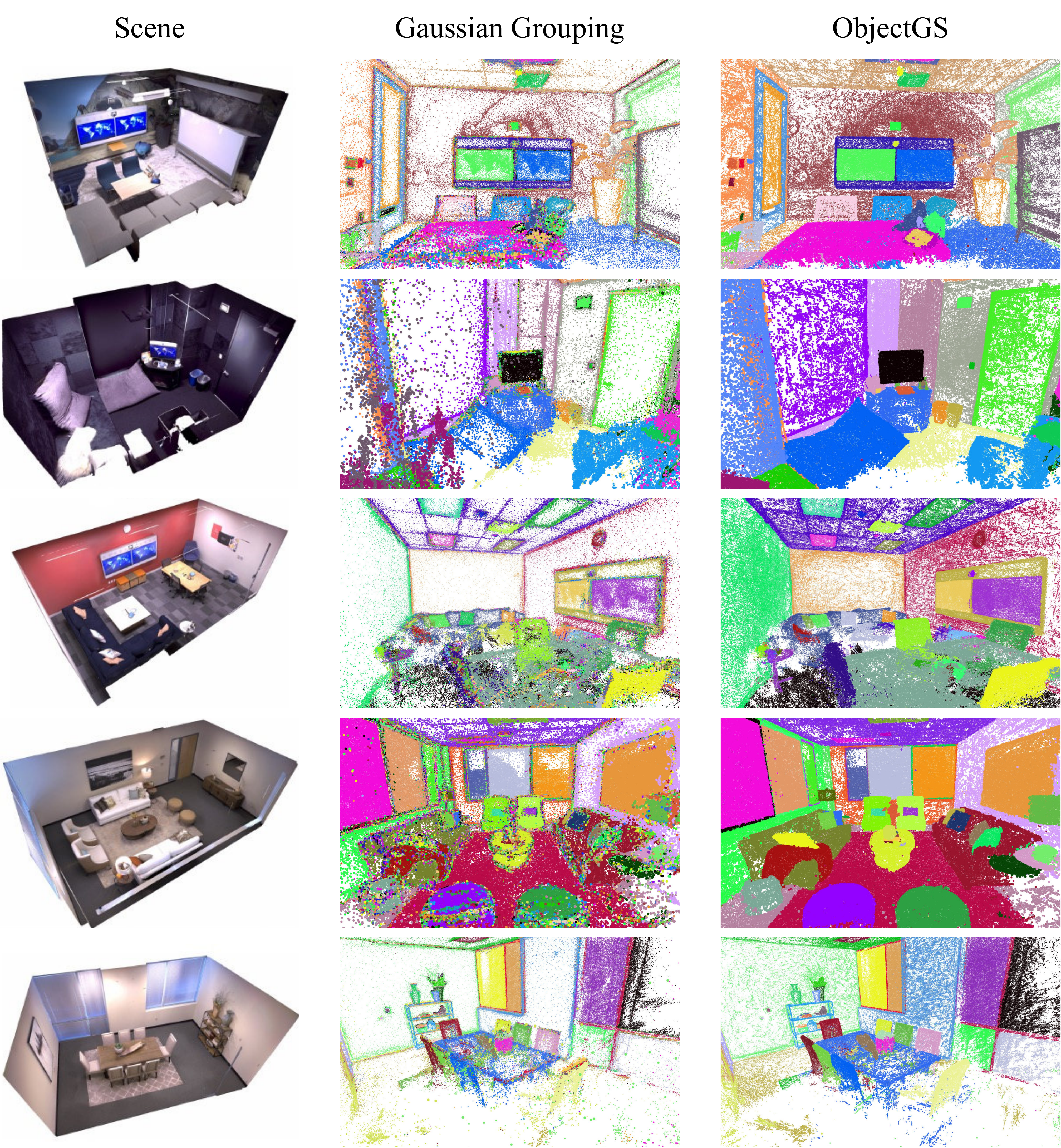}
    \caption{Qualitative comparison of 3D panoptic segmentation on the Replica dataset.}
    \label{fig:replica_pc}
\end{figure*}

\begin{figure*}
    \centering
    \includegraphics[width=\linewidth]{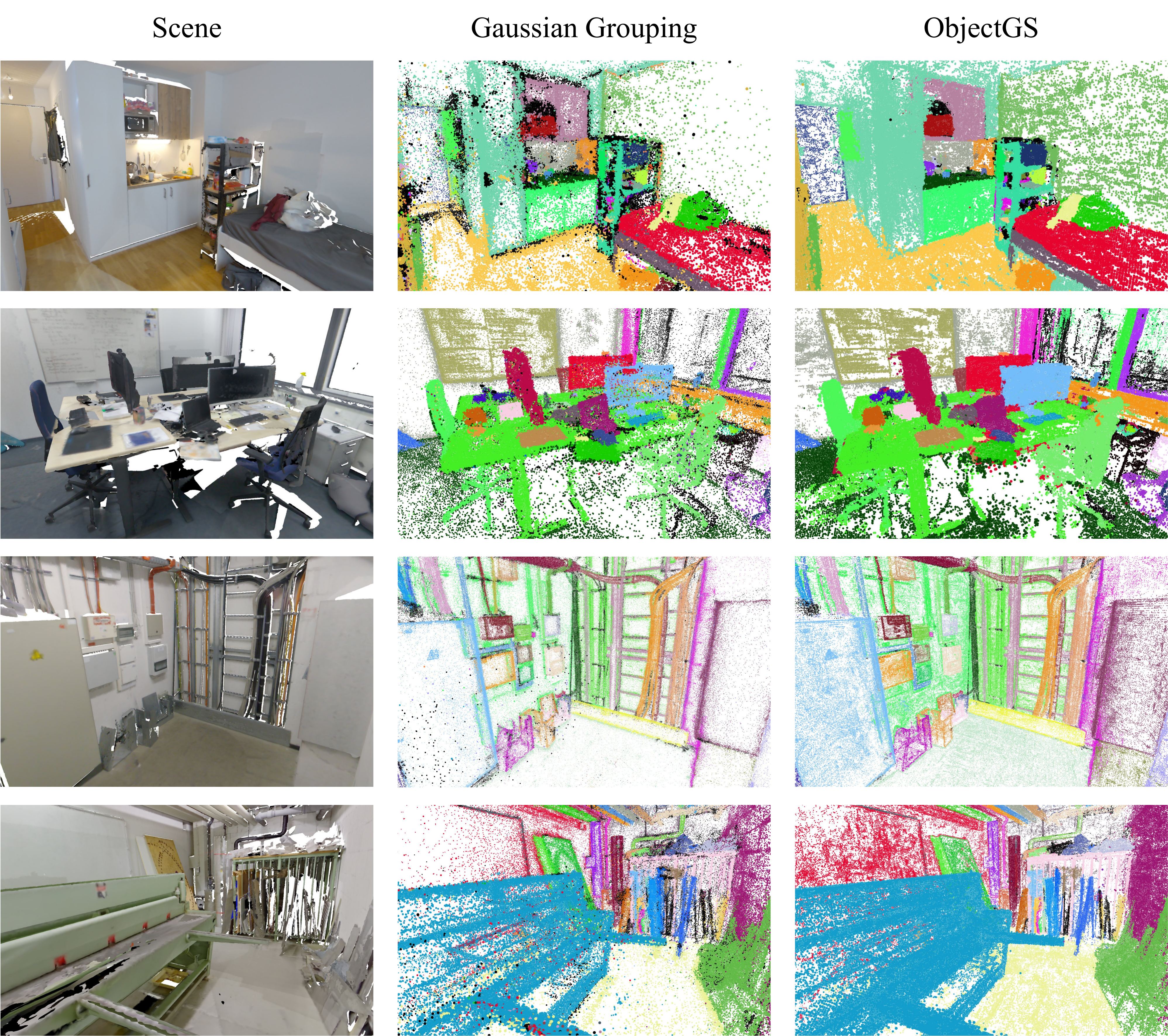}
    \caption{Qualitative comparison of 3D panoptic segmentation on the Scannet++ dataset.}
    \label{fig:scannetpp_pc}
\end{figure*}

\begin{figure*}
    \centering
    \includegraphics[width=\linewidth]{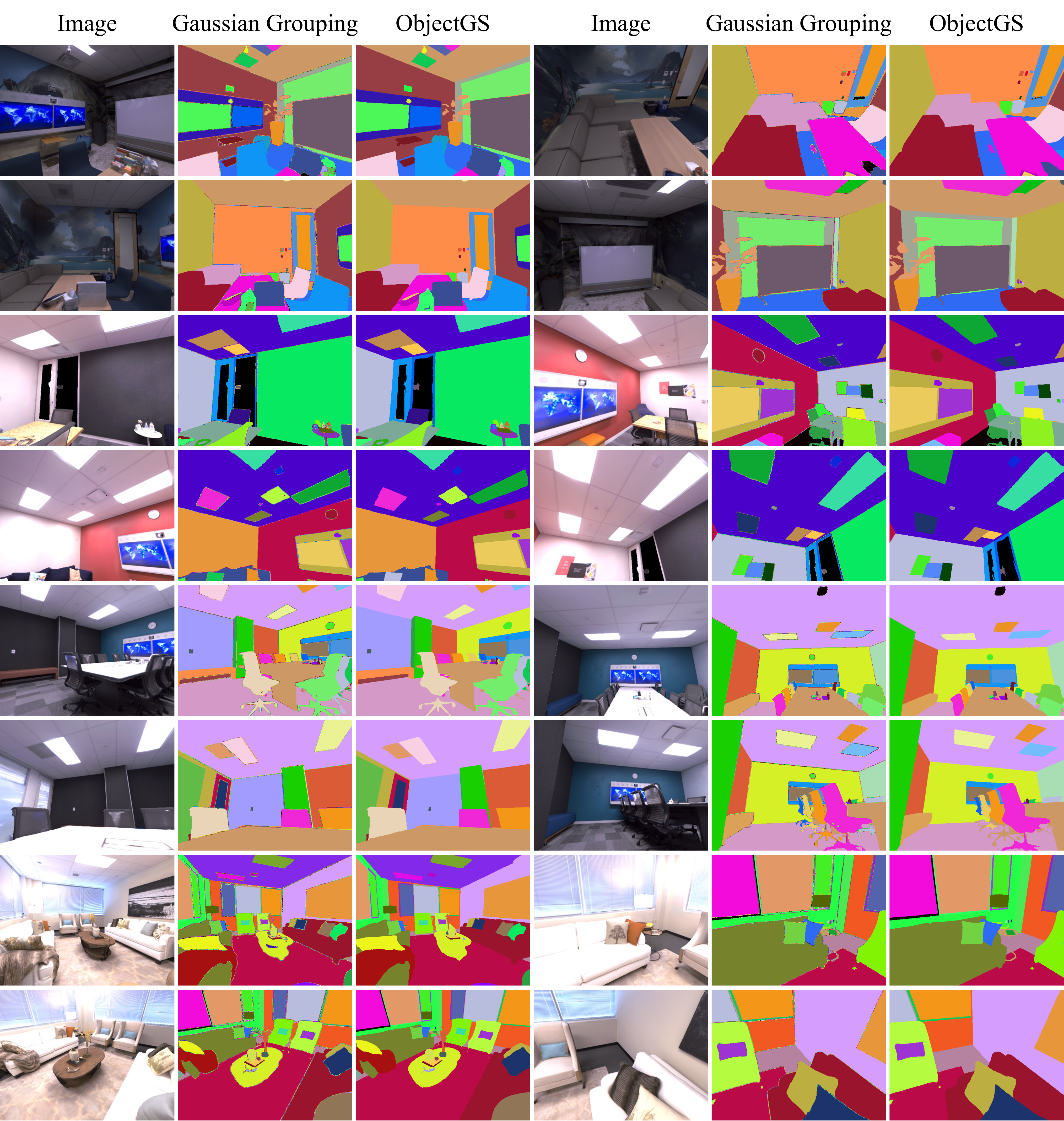}
    \caption{Qualitative comparison of 2D panoptic segmentation on the Replica dataset.}
    \label{fig:replica}
\end{figure*}

\begin{figure*}
    \centering
    \includegraphics[width=\linewidth]{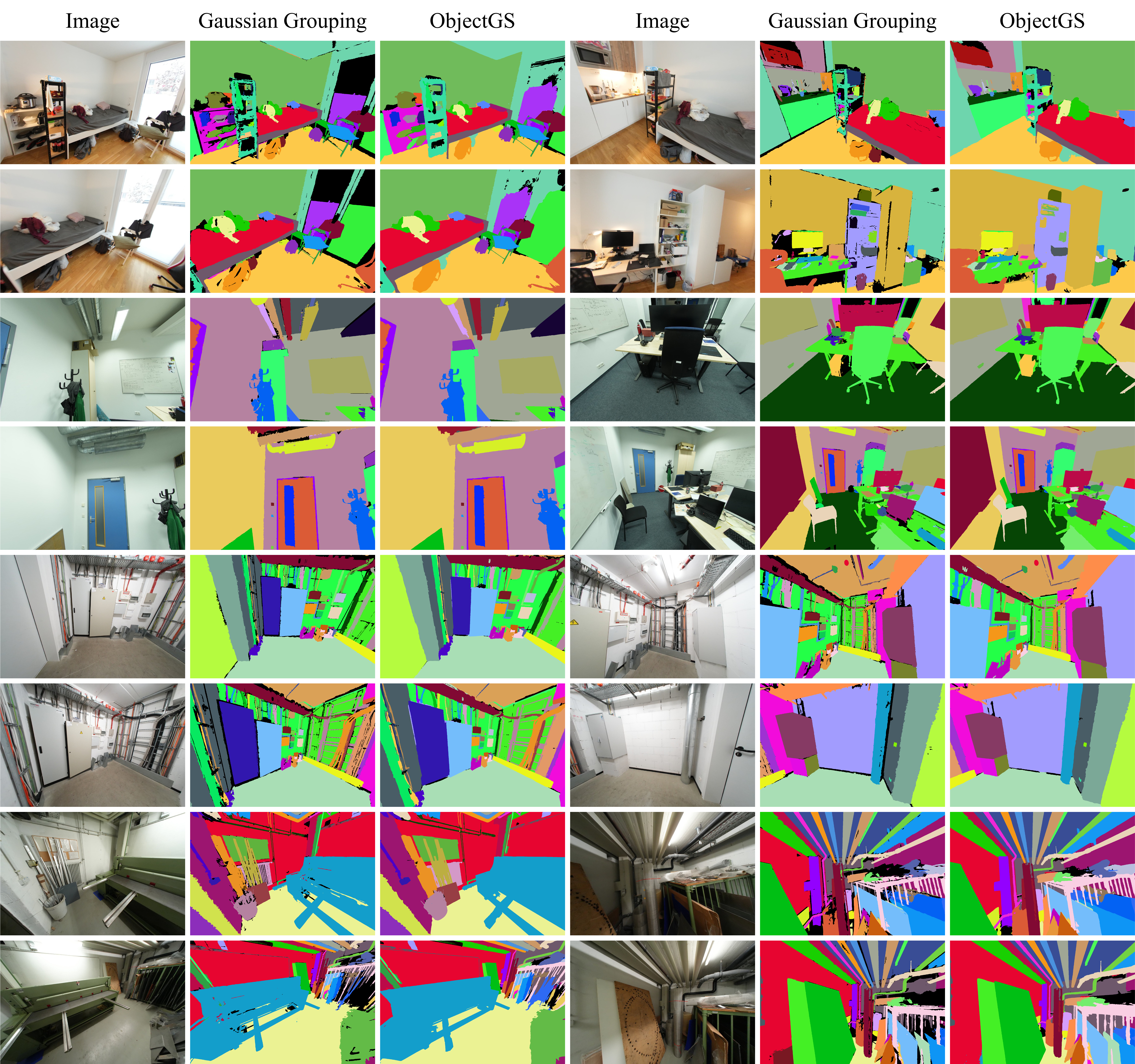}
    \caption{Qualitative comparison of 2D panoptic segmentation on the Scannet++ dataset.}
    \label{fig:scannetpp}
\end{figure*}

We provide more visualization results as shown in~\cref{fig:3dovs,fig:lerf-mask,fig:replica_pc,fig:scannetpp_pc,fig:replica,fig:scannetpp}, which includes visualization of OVS segmentation results, panoptic segmentation results, and 3D instance segmentation with point clouds. 

\begin{algorithm}
\caption{Object ID Majority Voting}
\label{alg:major_voting}
\begin{algorithmic}[1]
\State \textbf{Input:} 
\State \quad Point cloud: $ P_{\text{3D}} = \{ p_1, p_2, \dots, p_M \} $
\State \quad Object ID maps: $ L = \{ L_{1}, L_{2}, \dots, L_{N} \} $  
\State \quad Camera poses: $ C = \{ C_{1}, C_{2}, \dots, C_{N} \} $  

\State \textbf{Initialization:}
\State \quad $ \text{labels} = \emptyset $

\For{each point $ p_i \in P_{\text{3D}} $}
    \For{each camera pose $C_j \in C$}
        \State $x_i = \text{Project}(p_i, C_j)$
        \State Append $ L_{j}(x_i) $ to $ \text{labels}[p_i] $
    \EndFor
\EndFor

\For{each point $ p_i \in P_{\text{3D}} $}
    \If{ $ \text{labels}[p_i] \neq \emptyset $}
        \State $ \text{frequency(ID)} = \text{Counter}(\text{labels}[p_i]) $
        \State $ \text{ID} = \arg\max \text{frequency(ID)} $
    \EndIf
    \State Update $ p_i = (x_i, y_i, z_i, \text{object ID}) $
\EndFor

\State \textbf{Output:} Updated point cloud $ P_{\text{3D}} $ with object IDs.
\end{algorithmic}
\end{algorithm}

\begin{algorithm}
\caption{Object ID Probability-based Voting}
\label{alg:prob_voting}
\begin{algorithmic}[1]
\State \textbf{Input:} 
\State \quad Point cloud: $ P_{\text{3D}} = \{ p_1, p_2, \dots, p_M \} $
\State \quad Object ID maps: $ L = \{ L_{1}, L_{2}, \dots, L_{N} \} $  
\State \quad Camera poses: $ C = \{ C_{1}, C_{2}, \dots, C_{N} \} $  

\State \textbf{Initialization:}
\State \quad $ \text{labels} = \emptyset $

\For{each point $ p_i \in P_{\text{3D}} $}
    \For{each camera pose $C_j \in C$}
        \State $x_i = \text{Project}(p_i, C_j)$
        \State Append $ L_{j}(x_i) $ to $ \text{labels}[p_i] $
    \EndFor
\EndFor

\For{each point $ p_i \in P_{\text{3D}} $}
    \If{ $ \text{labels}[p_i] \neq \emptyset $}
        \State $ \text{frequency(ID)} = \text{Counter}(\text{labels}[p_i]) $
        \State $ \text{ID} = \text{Random} (\text{Prob} = \text{frequency(ID)}) $
    \EndIf
    \State Update $ p_i = (x_i, y_i, z_i, \text{object ID}) $
\EndFor

\State \textbf{Output:} Updated point cloud $ P_{\text{3D}} $ with object IDs.
\end{algorithmic}
\end{algorithm}

\begin{algorithm}
\caption{Object ID Correspondence-based Voting}
\label{alg:corr_voting}
\begin{algorithmic}[1]
\State \textbf{Input:} 
\State \quad Point cloud: $ P_{\text{3D}} = \{ p_1, p_2, \dots, p_M \} $
\State \quad Object ID maps: $ L = \{ L_{1}, L_{2}, \dots, L_{N} \} $  
\State \quad Correspondences: $ C = \{ C_{1}, C_{2}, \dots, C_{N} \} $  

\State \textbf{Initialization:}
\State \quad $ \text{labels} = \emptyset $

\For{each point $ p_i \in P_{\text{3D}} $}
    \For{each correspondence $C_j \in C$}
        \State $x_i = \text{Project}(p_i, C_j)$
        \State Append $ L_{j}(x_i) $ to $ \text{labels}[p_i] $
    \EndFor
\EndFor

\For{each point $ p_i \in P_{\text{3D}} $}
    \If{ $ \text{labels}[p_i] \neq \emptyset $}
        \State $ \text{frequency(ID)} = \text{Counter}(\text{labels}[p_i]) $
        \State $ \text{ID} = \arg\max\text{frequency(ID)} $
    \EndIf
    \State Update $ p_i = (x_i, y_i, z_i, \text{object ID}) $
\EndFor

\State \textbf{Output:} Updated point cloud $ P_{\text{3D}} $ with object IDs.
\end{algorithmic}
\end{algorithm}

\end{document}